\documentclass[aip,pop,reprint]{revtex4-1}
\usepackage[]{graphicx}
\usepackage[]{color}

\newcommand{\bra}[1]{\left<#1\right|}
\newcommand{\ket}[1]{\left|#1\right>}
\newcommand{\braket}[2]{\left<#1\middle|#2\right>}
\definecolor{comment}{rgb}{.94,0,.16}

\newcommand{\FEFF}{\texttt{FEFF}}
\newcommand{\pwffacites}{\cite{Riley2007,Sawada2007,Sahoo2008,Glenzer2009,Kritcher2009,Kritcher2011,Fortmann2012}}
\newcommand{\corecites}{\cite{Schuelke,Bradley2011,Bradley2010,Bradley2010b,Sakko2010,Feroughi2010,Sternemann2008,Fister2008,Gordon2008,Balasubramanian2007,Fister2006,Feng2004,Lee2008,Bergmann2004}}
\newcommand{\valencecites}{\cite{Cooper,Huotari2007,Volmer2007,Huotari2010,Huotari2009,Hakala2004,Okada2012,Sakurai2011}}

\begin{document}

%\title{Correct and Incorrect Treatments of the Bound-Free Contribution to X-ray Thomson Scattering}
\title{Theoretical Treatments of the Bound-Free Contribution and Experimental Best
Practice in X-ray Thomson Scattering from Warm Dense Matter}
\author{Brian A. Mattern}
\author{Gerald T. Seidler}
\email{seidler@uw.edu}
\affiliation{Department of Physics, University of Washington, Seattle, WA 98195-1560}
\date{\today}

\begin{abstract}
By comparison with high-resolution synchrotron x-ray experimental results,
we assess several theoretical treatments for the bound-free (core-electron)
contribution to x-ray Thomson scattering (i.e., also known as nonresonant
inelastic x-ray scattering). We identify an often overlooked source of
systematic error in the plane-wave form factor approximation (PWFFA) used
in the inference of temperature, ionization state, and free electron
density in some laser-driven compression studies of warm dense matter.
This error is due to a direct violation of energy conservation in the
PWFFA.\@  We propose an improved practice for the bound-free term that will
be particularly relevant for XRTS experiments performed with somewhat
improved energy resolution at the National Ignition Facility or the Linac
Coherent Light Source.  Our results raise important questions about the
accuracy of state variable determination in XRTS studies, given that the
limited information content in low-resolution XRTS spectra does not
strongly constrain the models of electronic structure being used to fit the
spectra.
\end{abstract}

\maketitle
\section{Introduction}
\label{sect:intro}

The field of \textit{warm dense matter} (WDM) rests in the transitional regime
between traditional condensed phase systems and strongly-correlated,
fully-ionized plasmas. As such, it draws from the complexity of both fields
while showing its own special fundamental and, as we show here, pragmatic
challenges.  A significant and growing literature exists on the electronic
structure,\cite{Gregori2003,Gregori2006,Trickey2011,Trickey2012,Plagemann2012}
thermodynamics\cite{Gregori2008} and hydrodynamics\cite{MacFarlane2006} of WDM,
in addition to a range of applications in fusion energy science\cite{Lindl2004}
and laboratory astrophysics.\cite{Remington2006,Wilson2012b} Here, however, we
investigate a particular difficulty of the WDM regime: assessing the accuracy
of the experimental determination of the basic state variables of the system,
such as temperature, density and ionization state.  Reliable inference of these
quantities is central to the clearly-needed improvements in the equation of
state of materials under, for example, the entire range of conditions leading
from ambient matter to inertial confinement fusion.\cite{Moses2009}

The high optical opacity of WDM requires the use of penetrating probe
radiation, i.e., x-ray photons with energies of a few to a few tens of keV.
Unfortunately, with only few counterexamples\cite{Doeppner2009} the inference
of state-variables using these methods is limited by the degree of
understanding of the electronic structure of WDM and its relationship to the
state variables themselves.  Faintly circular co-dependencies of this type are
not uncommon in emergent fields of experimental science (e.g., consider the
many years of effort needed to establish accurate and precise pressure sensing
in the Mbar range in opposed-anvil pressure
cells\cite{Forman1972,Piermarini1975,Gupta1990,Ragan1992}), and a firm
foundation for such methodologies can follow from any of several developments.
Foremost among such developments are: experimental data of sufficient
information content to itself strongly constrain the constituent theories for
electronic structure; broad programs to assess accuracy by cross-comparison of
different metrologies; and, finally, an eventual comparison to international
standards.  The experimental determination of state variables in the WDM regime
is seeing only the earliest such examples, including notably the rare
experiments using detailed balance in x-ray Thomson
scattering\cite{Doeppner2009} or the recent checks in consistency between
conclusions drawn from the elastic and inelastic components of x-ray
scattering.\cite{Fortmann2012}

The present paper reports a first step in evaluating the accuracy, rather than
precision, of the methods used for state variable determination in the WDM
regime.  To this end, we investigate the various available treatments for the
core-electron, or bound-free, contribution to x-ray Thomson scattering (XRTS,
also called nonresonant inelastic x-ray scattering, NIXS\cite{Schuelke}) with a
special emphasis on the plane-wave form-factor approximation (PWFFA) of
Schumacher, et al.\cite{Schumacher1975}. We will show that this approximation,
which has seen extensive use in XRTS studies of shock-compressed
matter,\pwffacites is fundamentally flawed and presents a source of systematic
uncertainty in inferred quantities.

Based on these observations we come to three main conclusions.  First, looking
to the near future, when XRTS studies of WDM with improved energy resolution
will be performed at the Linac Coherent Light Source\cite{HauRiege2012} and
the National Ignition Facility\cite{Moses2009}, the errors implicit in the use
of the PWFFA must be avoided if physically meaningful information on the
equation of state in the WDM regime is to be determined.  Second, while it is
important to note that a faulty theoretical treatment has been used, and that
some reevaluation of experimental results may be called for, the more lasting
conclusion is that the information content in the measured XRTS spectra for
WDM has been insufficient to alert the experimenters to the presence of an
unphysical model for the electronic structure.  This strongly suggests the
need for cross-comparison with alternative methods of WDM state variable
determination, e.g., x-ray fluorescence
thermometry\cite{Zastrau2010,Sengebusch2009,Stambulchik2009,Nilson2010,Levy2012,Vinko2012}.
Third, we find that stronger connections between the synchrotron x-ray and
WDM-XRTS communities provide important experimental and theoretical synergies.
The wealth of very high-resolution studies at synchrotron light sources of
both the free-free (valence)\valencecites and bound-free (core) \corecites
contributions to XRTS provide important benchmarks both for comparison to
WDM-specific theory and also for validation of experimental protocol
including, e.g., instrument-specific backgrounds.  Further, as we have
illustrated here, there will be cases where theoretical methods already in use
for synchrotron studies may be beneficially transported to WDM-XRTS studies.

In Sect.~\ref{sect:theory} we discuss four theoretical treatments of bound-free
XRTS: the impulse approximation (IA), which is valid at large energy transfer;
a hydrogenic model (HM); an extension of the IA to incorporate binding energies
(PWFFA); and a real-space Green's function method (RSGF). The first three are
essentially atomic (with varying degrees of approximation), while the latter
treats the condensed solid, and is based on methods broadly used for many years
in the interpretation of several x-ray spectroscopic techniques.
\cite{RehrAlbers2000,Rehr2009}

Since high-resolution XRTS data from WDM at known thermodynamic conditions is
currently unavailable, we instead compare each of the above theories with very
high-quality XRTS spectra collected under ambient conditions at a synchrotron x-ray
source. This provides a baseline validation for the core contribution: a
theoretical treatment which fails under these conditions is certain to form a
weak foundation when including the further complexities of continuum lowering and
partial ionization present in WDM.

Experimental details are described in Sect.~\ref{sect:experiment} and the
comparison of theory and experiment is made in Sect.~\ref{sect:comparison}.
The relative success of even atomic treatments at describing the condensed
solid suggests that these methods should be extensible to the WDM regime with
only minor modifications.  However, we find that the PWFFA, which has been used
for a few years in the interpretation of WDM measurements,\pwffacites is in
stark disagreement with the ambient experimental data.  In
Sect.~\ref{sect:pwffa_in_detail}, we show that this disagreement is due to
internal inconsistency in the PWFFA that leads to unphysical results.  Next, in
Sect.~\ref{sect:implications}, we consider the implications of using the PWFFA
to model the bound-free contribution to WDM XRTS data; namely, the likelihood
of previously-undiagnosed systematic errors in extracted thermodynamic
quantities.  These observations then motivate a discussion of best future
practice in Sect.~\ref{sect:future_practice}, after which we conclude in
Sect.~\ref{sect:conclusions}.

\section{Theory}
\label{sect:theory}

The fundamental observable in XRTS/NIXS is the \textit{dynamic structure
factor} $S(\vec{q},\omega)$, which separates into independent contributions
from electrons in different shells.  We will focus on the contribution from
tightly bound core electrons, i.e., the \textit{bound-free} contribution.
The theoretical description of bound-free XRTS begins with the
Kramers-Heisenberg formula for the first-Born approximation to the
double-differential scattering cross-section (DDSCS):\cite{Schuelke}
\begin{eqnarray}
  \label{eq:DDSCS}
  \frac{d^2\sigma}{d\omega d\Omega} &=& \frac{\omega_2}{\omega_1} r_o^2 \left|\hat{\epsilon}_1 \cdot \hat{\epsilon}_2^*\right|^2 S(\vec{q}, \omega) \\
  S(\vec{q}, \omega) &=& \sum_I P_I(T) \sum_F \Big| \bra{F} \sum_j e^{i\vec{q}\cdot\vec{r}_j} \ket{I} \Big|^2 \nonumber \\
  &\times& \delta(E_F - E_I - \omega)
  \label{eq:Sfull}
\end{eqnarray}
Here, $\omega_{1,2}$ and $\hat{\epsilon}_{1,2}$ are initial and final photon
energies and polarizations; $r_0$ is the Thomson scattering length;
$\ket{I}$,$\ket{F}$ are initial and final many-body states with energies
$E_I$,$E_F$; $P_I(T)$ is the temperature-dependent Boltzmann factor; $\vec{q}$
is the momentum transfer; $\omega = \omega_1 - \omega_2$ is the energy
transfer and $j$ indexes individual electrons.  In this and subsequent
formulae, we use Hartree atomic units ($\hbar = m_e = 1$).

In the independent particle approximation, Eq.~(\ref{eq:Sfull}) can written as
\begin{eqnarray}
  S(\vec{q}, \omega) &=& \sum_i n_i S_i(\vec{q}, \omega)
  \nonumber \\ 
  \label{eq:Si}
  S_i(\vec{q}, \omega) &=& \sum_f (1-n_f) \left| \bra{f} e^{i\vec{q}\cdot\vec{r}} \ket{i} \right|^2 \delta(E_f - E_i - \omega),
\end{eqnarray}
where $\ket{i}$,$\ket{f}$ are initial and final state \textit{single particle}
states with energies $E_i$,$E_f$ and thermally-averaged occupation numbers $n_i$, $n_f$.

\subsection{The impulse approximation}
In the limit of large energy-transfer $\omega$ relative to the initial state
binding energy $E_B$, known as the impulse approximation (IA), the XRTS spectrum
is completely determined by the
initial-state electronic momentum distribution;  the binding energy of the scattering
electron plays no
role.\cite{Eisenberger1970}

For $\omega \gg E_B$, only unoccupied final states contribution to
Eq.~(\ref{eq:Si}), so we set $n_f = 0$.  Next, following Eisenberger and
Platzman,\cite{Eisenberger1970} we expand the $\delta$-function using the
standard Fourier representation
\begin{equation}
  \delta(\omega) = \int \frac{dt}{2\pi} e^{i\omega t}.
  \label{eq:deltaFourier}
\end{equation}
After rearranging slightly and using the fact that $\ket{i}$ and $\ket{f}$ are
eigenstates of the single-particle Hamiltonian $H$, we find
\begin{eqnarray}
  S_i(\vec{q}, \omega) &=& \int \frac{dt}{2\pi} \sum_f e^{i\omega t} \bra{i}e^{iHt} e^{-i\vec{q}\cdot\vec{r}} e^{-iHt}\ket{f} \bra{f}e^{i\vec{q}\cdot\vec{r}}\ket{i}  \nonumber\\
  &=& \int \frac{dt}{2\pi} e^{i\omega t} \bra{i}e^{iHt} e^{-i\vec{q}\cdot\vec{r}} e^{-iHt}e^{i\vec{q}\cdot\vec{r}}\ket{i}.
  \label{eq:IA_pre_approx}
\end{eqnarray}
In the second line, we have used completeness to remove the sum over final
states.  The IA corresponds to replacing $H$ by the free-particle Hamiltonian
$H_0$, which can be justified in the limit of $(\omega/E_B)^2 \gg 1$ (see
Section III of Ref.~\onlinecite{Eisenberger1970}). After inserting a complete
set of momentum eigenstates and integrating over the direction of momentum, we
obtain
\begin{equation}
  S_i(\vec{q}, \omega) = (2\pi/q)\int_{|w/q - q/2|}^{\infty} p\, dp\, \rho_i(p),
  \label{eq:IA}
\end{equation}
where $\rho_i(p) = (2\pi)^{-3}|\braket{i}{p}|^2$ is the initial-state momentum
density, which is here assumed to be isotropic (e.g.\ $s$-shell, or sum over a
filled subshell).

This formula can be interpreted as describing XRTS from a gas of free electrons
with the same initial-state momentum distribution.  The scattering spectrum
consists of a line centered at the free-particle Compton shift $\omega_c =
q^2/2$ that is Doppler broadened by the projection of the momentum distribution
along the direction of $\vec{q}$.  The integral over the momentum distribution
in Eq.~(\ref{eq:IA}) simply counts electrons with a given momentum-projection
$p_q$ determined by the energy transfer.

We now turn to methods that take the binding energy into account.

\subsection{Hydrogenic Model}

It is possibly to analytically evaluate Eq. (\ref{eq:Si}) using hydrogenic initial and final states. For a $1s$ initial state, the result is\cite{Eisenberger1970}
\begin{equation}
  S_i(q,\omega) = \int \frac{d^3p}{(2\pi)^3} |\bra{f}e^{i\vec{q}\cdot\vec{r}}\ket{i}|^2 \delta(\omega - E_B - p^2/2),
  \label{eq:hydrogenic1}
\end{equation}
with
\begin{eqnarray}
  |\bra{f}e^{i\vec{q}\cdot\vec{r}}\ket{i}|^2 &=& 
  \frac{\pi^28^3a^2}{p}(1 - e^{-2\pi/pa})^{-1} \nonumber\\
  &\times& \exp\left[\frac{-2}{pa}\tan^{-1}\left(\frac{2pa}{1+(q^2a^2)-p^2a^2}\right)\right] \nonumber \\
  &\times& \left[ k^4a^4 + (1/3)k^2a^2(1+p^2a^2) \right] \nonumber\\
  &\times& \left[(k^2a^2 + 1 - p^2a^2)^2 + 4 p^2a^2\right]^{-3}.
  \label{eq:hydrogenic}
\end{eqnarray}
Here, $p$ is the final-state momentum, $a = 1/Z$ where $Z$ is the effective
nuclear charge\cite{Clementi1963}, and $E_B = 1/2Z^2$ is the binding energy.  
In this expression, contributions from bound final states have been neglected.
Expressions for shells other than the $1s$ are included in Schumacher, et al.\cite{Schumacher1975}
, where this method is referred to as the \textit{hydrogenic form-factor approximation}.
 We will, however, refer to this simply as the hydrogenic model (HM).

\subsection{Plane-wave form-factor approximation}
\label{sect:pwffa}

The \textit{plane-wave form-factor approximation} (PWFFA) is an attempt to
improve the IA by including the binding energy in the kinematics. The final
states are assumed to be momentum eigenstates, which is conceptually appealing
in the context of dense plasmas where the jellium model has found wide
application.  However, as we demonstrate in Sect.~\ref{sect:pwffa_in_detail},
a fundamental difficulty arises: this approximation effectively evaluates the
initial-state energy using the atomic Hamiltonian, while evaluating the
final-state energy using the free-particle Hamiltonian. This inconsistent
treatment violates energy conservation, resulting in violations of the Bethe
$f$-sum rule and deviations from experimental results that, although small in
the original context of gamma-ray scattering, are quite large under the
kinematic conditions typical of XRTS measurements.  It is the use of the PWFFA
in the interpretation of recent XRTS experiments on WDM\pwffacites that
motivates the present paper.

Schumacher's derivation of the PWFFA\cite{Schumacher1975} makes the
following assumptions:
\begin{eqnarray}
  \ket{f} &=& \ket{\vec{p}+\vec{q}} \nonumber\\
  \sum_f &\rightarrow& \int  \frac{d^3p }{(2\pi)^3} \nonumber\\
  E_f &=& E_{\vec{p}+\vec{q}} = \frac{(\vec{p}+\vec{q})^2}{2} \nonumber\\
  E_i &=& -E_B
  \label{eq:assumptions}
\end{eqnarray}
where $E_B$ is the initial-state binding energy, $\vec{p}$ is the initial-state
momentum  and $\vec{p}+\vec{q}$ is the final-state momentum. Assuming that $T=0$
and applying (\ref{eq:assumptions}) to (\ref{eq:Si}), we find
\begin{eqnarray}
  S_i(\vec{q}, \omega) &=& \int \frac{d^3p}{(2\pi)^3}  \left| \bra{\vec{p} + \vec{q}} e^{i\vec{q}\cdot\vec{r}} \ket{i} \right|^2 \delta(E_{\vec{p}+\vec{q}} + E_B - \omega) \nonumber\\
  &=& \int \frac{d^3p}{(2\pi)^3}  \left| \braket{\vec{p}}{i} \right|^2 \delta(E_{\vec{p}+\vec{q}} + E_B - \omega) \nonumber \\
  &=& \int d^3p \rho_i(\vec{p}) \delta(E_{\vec{p}+\vec{q}} + E_B - \omega)
  \label{eq:derivation}
\end{eqnarray}
In the second line we use the fact that $e^{i\vec{q}\cdot\vec{r}}$ is a
momentum translation operator. The last line uses the definition of the
momentum density $\rho_i(\vec{p}) =
(2\pi)^{-3}\left|\braket{\vec{p}}{i}\right|^2$. Furthermore, if we again
restrict ourselves to an isotropic momentum density (e.g., $s$-shell, or sum
over a filled subshell), we can perform the angular integrals to obtain
\begin{equation}
  S_i(\vec{q}, \omega) = (2\pi/q) \int_{|\sqrt{2(\omega-B)}-q|}^{\sqrt{2(\omega-B)}+q} p\, dp\, \rho_i(p)
  \label{eq:PWFFA}
\end{equation}
This expression, along with (\ref{eq:DDSCS}) differs from
Schumacher's\cite{Schumacher1975} Eqs. (5) and (21) only in that it does not
contain the relativistic prefactor $\sqrt{1+(p/mc)^2}$, which for the
experimental conditions under consideration differs negligibly from unity.

Eq.~(\ref{eq:PWFFA}) has the same form as the IA expression Eq.~(\ref{eq:IA}),
differing only by the bounds on the integration over the momentum density.
Given this similarity and the well-tested validity of the IA in its regime of
applicability, one would expect that for $\omega \gg E_B$ the PWFFA would
reproduce the IA\@. This is, however, not the case --- despite claims in the
literature to the contrary.\cite{Schumacher1975,Glenzer2009} We will return to
this point in Sect.~\ref{sect:results}, after comparison with experiment.

\subsection{The real-space Green's function method}

The prior methods have all treated XRTS for an isolated atom to various
degrees of approximation.  The next method we describe treats an arbitrary
cluster of atoms using a real-space Green's function (RSGF) formalism
implemented in recent versions of the x-ray spectroscopy code
\FEFF.\cite{Soininen2005}  This formalism, which can treat complex, aperiodic
systems, has been extensively applied to condensed-matter systems, where
XRTS/NIXS provides a bulk-sensitive alternative to, and extension of, soft
x-ray absorption spectroscopy,
\cite{Soininen2005,Sternemann2007,Sternemann2007b,Balasubramanian2007,Feng2008,Fister2009,Pylkkanen2010}
The RSGF approach has recently been extended to treat the valence
contribution.\cite{Mattern2012}

Starting with a description of atomic species and locations, an effective
one-particle Green's function for the valence electrons in the cluster of atoms
is calculated in the muffin-tin approximation, including the effects of full
multiple scattering\cite{Rehr2009}.  This Green's function implicitly contains the excited
electronic states that are the final states in the scattering experiment. In
terms of a spectral density matrix defined by $\rho(E) = \sum_f
\ket{f}\bra{f}\delta(E-E_f)$, which is related to the Green's function by
$\bra{\vec{r}}\rho(E)\ket{\vec{r}'} = -
(1/\pi)\,\textrm{Im}\,G(\vec{r}',\vec{r},E)$, Eq. (\ref{eq:Si}) can be recast
as
\begin{equation}
  S_i(\vec{q}, \omega) = \bra{i}e^{-i\vec{q}\cdot\vec{r'}}P\rho(E) P e^{i\vec{q}\cdot\vec{r}}\ket{i}.
  \label{eq:Srsms}
\end{equation}
Here $E = \omega + E_i$ is the photoelectron energy and $P$ projects the final
states (which are calculated in the presence of a core hole) onto the
unoccupied states of the initial-state Hamiltonian (which has no core
hole).\cite{Soininen2005} The Green's function can be separated into
contributions from the central atom and from scattering off other atoms in the
cluster. Likewise, the dynamic structure factor can be factored as 
\begin{equation}
  S_i(\vec{q}, \omega) = S_0(q, \omega)[1 + \chi_{\vec{q}}(\omega)],
\end{equation}
where $S_0(q,\omega)$ is a smoothly varying, isotropic atomic background and
$\chi_{\vec{q}}$ is the fine structure due to all orders of photoelectron
scattering from the environment.\cite{Soininen2005,Fister2006} Implicit in the
fine structure is information about nearest-neighbor distances and thus also
density\cite{Soininen2005}. However, at the poor experimental resolution
typical of WDM measurements\cite{Glenzer2009}, this structure will be washed
out.  Thus, for our purposes, we will include only the atomic background
contribution, $S_0(q, \omega)$.

\section{Experiment}
\label{sect:experiment}

\begin{figure}
  \begin{center}
    \includegraphics{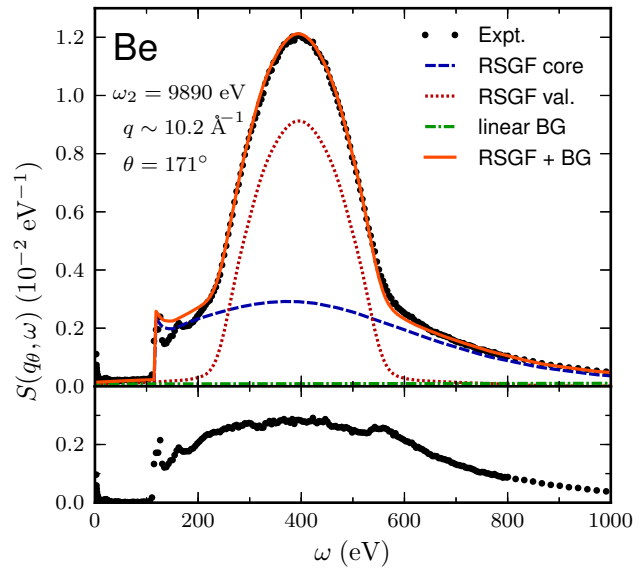}
  \end{center}
  \caption{(Color online.) XRTS from polycrystalline beryllium under ambient
    conditions at a fixed $171^\circ$ scattering angle and 9890-eV scattered
    photon energy\cite{Mattern2012}.  Here, the energy transfer $\omega$ is
    the difference between the incident and scattered photon energies.  In the
    upper panel, the data are shown along with a combined real-space Green's
    function (RSGF) valence and core calculation.  The data have been scaled
    as described in the text. In the lower panel, the valence contribution and
    linear background have been subtracted to give the core contribution alone.}
  \label{fig:expt}
\end{figure}

Although we are ultimately interested in the elevated temperatures and
densities of WDM, it is important to first validate theoretical methodology
against spectra taken under known thermodynamic conditions and at higher
resolution than presently typical of WDM experiments. To this end,
experimental XRTS data for polycrystalline Be at ambient temperature and
pressure were collected using the lower energy resolution inelastic x-ray
(LERIX) spectrometer at beamline 20-ID of the Advanced Photon
Source.\cite{LERIX} Scattered photons with $\omega_2 = (9891.7 \pm 0.2)$~eV
were analyzed by a single spherically bent Si crystal located at a fixed
$171^\circ$ scattering angle while scanning the incident photon energy. From
the elastic peak width the total instrumental resolution was determined to be
1.3 eV. The data, which have previously been reported\cite{Mattern2012}, are
shown in Fig.~\ref{fig:expt}. The graph is labelled $S(q_\theta, \omega)$ to
indicate that the data are collected at fixed scattering angle, and thus $q$
is a weak function of $\omega$.

After normalizing to the incident flux, a small linear background and a single
scale factor were fit in order to match the RSGF core calculation in the tail
region ($800 < \omega < 1500$ eV) and thus put the data into absolute units.
The results of this fit have been used to scale the experimental data in
Fig.~\ref{fig:expt}. Additionally, theoretical core and valence calculations,
the fit linear background, and the sum of these three are shown.

The experimental generalized oscillator strength $\int (2/q^2) \omega
S(q,\omega) d\omega$ matches that for the combined theoretical RSGF spectrum to
within 1\%. Since the measurement was performed at fixed scattering angle, this
value is slightly larger than the Bethe $f$-sum
rule\cite{Schuelke,Inokuti1978,Wang1999} value of N=4 (which only holds for
experiments performed at fixed $q$).

The theoretical valence profile, calculated from the RSGF,\cite{Mattern2012}
was then subtracted to obtain the experimental core profile, shown in the
lower panel of Fig.~\ref{fig:expt}.  The small peak visible for $550 < \omega <
650$ eV is a result of the theoretical valence profile underestimating the
actual contribution in this region, as was also seen in comparisons with higher
momentum-transfer data (see Figs. 4 and 5 of Mattern, et al.\cite{Mattern2012}).

\section{Results and Discussion}
\label{sect:results}

\begin{figure}[]
  \begin{center}
    \includegraphics{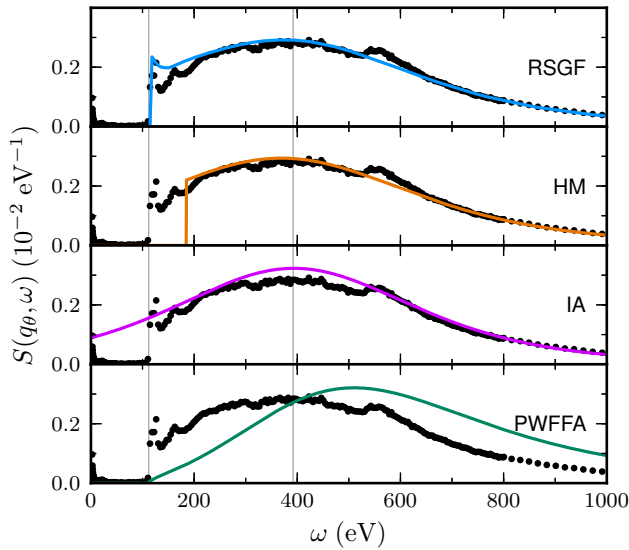}
  \end{center}
  \caption{(Color online.) Comparison of extracted core-shell XRTS with theoretical
    calculations using the RSGF, the hydrogenic model (HM), impulse
    approximation (IA), and plane-wave form-factor approximation (PWFFA).  The
    energy transfer $\omega$ is the difference between the incident and
    scattered photon energies.  \textit{All calculations are in absolute
    units.} Vertical guides are shown at the $1s$ binding energy (112 eV) and
    the free-particle Compton shift (396 eV).}
  \label{fig:be_theory}
\end{figure}

\begin{figure}
  \includegraphics{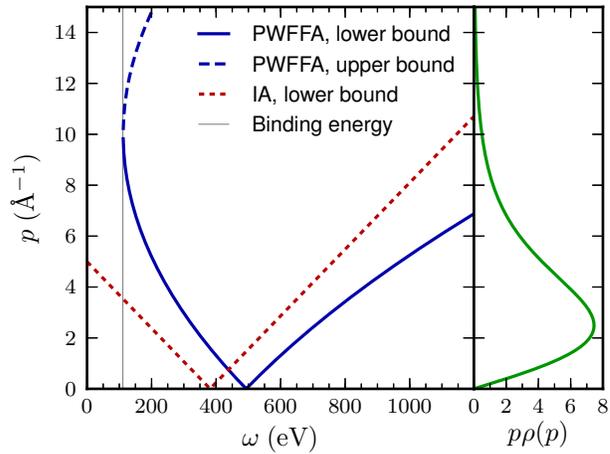}
  \caption{(Color online.) On the left are shown the integration bounds for both the PWFFA (upper and lower bounds) and IA (lower only, upper bound is $\infty$) calculations. On the right, plotted vertically, is the integrand $p\rho(p)$. Both the offset of the peak by the binding energy (112 eV) and the lack of convergence of the PWFFA to the IA at large $\omega$ are apparent.}
  \label{fig:bounds}
\end{figure}

\subsection{Comparison of Experiment and Theory}
\label{sect:comparison}

The extracted experimental core profile is compared in Fig.~\ref{fig:be_theory}
to to each of the four theoretical calculations discussed in
Section~\ref{sect:theory}. All calculations are in absolute units and have
taken the weak dependence of $q$ on $\omega$ into account.  Vertical guides are
included at the $1s$ binding energy (112 eV) and the free-particle Compton
shift (396 eV).

The IA and PWFFA calculations use the ground-state Dirac-Fock Be $1s$
wavefunction calculated using \FEFF's atomic solver as the initial state, and
thus only differ in their treatment of the final states and energy
conservation. For the RSGF calculation, the Dirac-Fock wavefunction is
calculated in the presence of a $1s$ core-hole. We have included only the
atomic background contribution $S_0(q, \omega)$. The fine structure visible in
the experimental data for $\omega \lesssim 200$ eV is not included, although it
has been treated elsewhere.\cite{Soininen2005}  The HM calculation uses
hydrogenic wavefunctions with an effective nuclear charge
(Z=3.685)\cite{Clementi1963} for the initial and final states.

We focus on three regions for comparison: the vicinities of the $K$-edge binding
energy ($\omega\sim 112$~eV), the peak ($\omega\sim 396$~eV), and the tail
($\omega \gtrsim 600$~eV). The RSGF calculation matches the data reasonably
well in all three regions (with the exception of immediately above the $K$ edge,
where interference effects have been omitted).
The HM accurately describes the peak and tail regions, but shows a large
deficit above the experimental binding energy due to the larger binding energy
of the hydrogenic state.  The IA, which ignores the binding energy, has an
unphysical tail at low energy transfers. The peak region is reasonably well
described by the IA, while the high-$\omega$ tail region is quite accurate.
This is expected since the conditions of applicability of the IA are well
satisfied for large energy transfer.

By contrast, although the PWFFA of Schumacher, et al.\cite{Schumacher1975}
vanishes below $K$ edge, it exhibits only a gradual onset and further shows
strong quantitative and qualitative disagreement with the experimental data
everywhere else.  Given its application in the interpretation of several XRTS
experiments\pwffacites on WDM, and its evident failure to describe
high-resolution synchrotron measurements,  we now turn our focus to the
PWFFA\@.  We will first look in more detail at the source of the
approximation's error, and then briefly discuss the possible implications for
interpretation of experiment and future best practice.

\subsection{A closer look at the PWFFA}
\label{sect:pwffa_in_detail}

Although others have previously observed that the PWFFA gives results with
unphysical features that are in disagreement with experimental
data,\cite{Currat1971,Bell1986} we are unaware of a discussion of the origin of
the approximation's inconsistency.  We now consider this point in detail.

An alternative route to obtain the PWFFA is to follow the IA
derivation up to Eq.  (\ref{eq:IA_pre_approx}). At this point, if one makes the
\textit{ad hoc} approximation of replacing only the second $H$ by $H_0$,
\begin{equation}
  S_i(\vec{q}, \omega) \approx \int \frac{dt}{2\pi} e^{i\omega t}
  \bra{i}e^{iHt} e^{-i\vec{q}\cdot\vec{r}}
  e^{-iH_0t}e^{i\vec{q}\cdot\vec{r}}\ket{i}.  \label{eq:PWFFA_flaw}
\end{equation}
then, instead, the PWFFA result (\ref{eq:PWFFA}) follows. Thus, the assumptions
(\ref{eq:assumptions}) correspond to making the uncontrolled approximation of
evaluating the initial-state energy using $H$ and the final-state energy using
$H_0$. This effectively violates energy conservation, opening the possibility
of unphysical results. 

As we mentioned in Sect.~\ref{sect:pwffa}, the IA~(\ref{eq:IA}) and
PWFFA~(\ref{eq:PWFFA}) results differ only in the bounds of the integration
over the momentum density. In left panel of Fig.~\ref{fig:bounds}, we show the
integration bounds as a function of $\omega$ for both theories. For the IA,
there is no upper bound, and for the PWFFA the upper bound is only relevant for
small $\omega$, beyond which it is well above the integrand's region of
support. The right panel shows the integrand (rotated so that the abscissa runs
vertically). The scattering spectra can be seen to peak when the lower
integration bound vanishes. For the PWFFA, this is offset to higher energy
transfer by the binding energy of the initial state. Furthermore, the failure of the
PWFFA to reduce to the IA at large $\omega$ can be clearly seen here.  The
offset of the PWFFA peak relative to the IA appears to be in conflict with the
calculations presented in Fig. 3 of Riley, et al.\cite{Riley2007}.

\begin{figure}
  \begin{center}
    \includegraphics{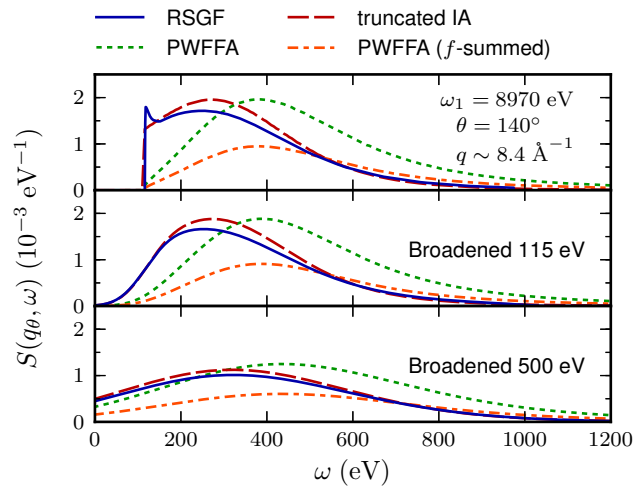}
  \end{center}
  \caption{(Color online.) Theoretical core profiles for the experimental conditions of Fortmann, et al.\cite{Fortmann2012}. The inaccuracy of the PWFFA remains, even after substantial broadening.}
  \label{fig:modelWDM}
\end{figure}

\begin{figure}
  \includegraphics{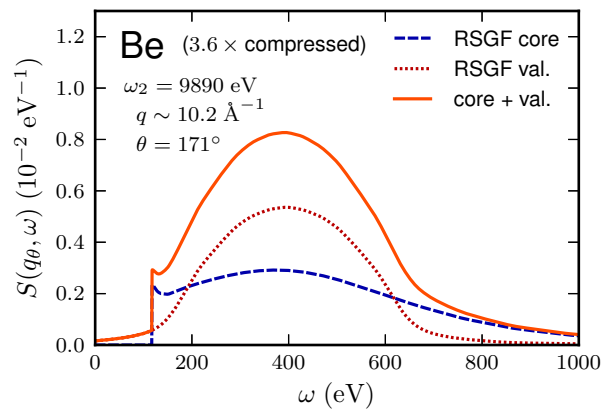}
  \caption{\
    (Color online.) Theoretical (RSGF) XRTS from polycrystalline Be compressed
    to $3.6\times$ ambient density.  The valence contribution is broader than
    under ambient conditions (\textit{cf.} Fig.~\ref{fig:expt}, top panel).
    Subsequently, the core contribution (assumed here to be independent of
    density) is relatively larger in the region of the peak.
  }
  \label{fig:compressed}
\end{figure}

Given the much lower energy resolution typical of WDM experiments, it is
natural to ask whether the discrepancies seen above are relevant in that
context.  In Fig.~\ref{fig:modelWDM}, we compare RSGF, IA and PWFFA
calculations at the representative experimental conditions of
Fortmann, et al.\cite{Fortmann2012}. The IA calculation has been truncated at the
empirical binding energy. In addition to the absolute-unit PWFFA calculation,
we have included a curve scaled to match the $f$-sum of the other theoretical
curves. The unbroadened calculations are shown in the upper panel. The curves
in the middle panel are broadened by 115-eV (FWHM) to match the experimental
resolution of Ref.~\onlinecite{Fortmann2012}. Finally, for illustrative
purposed, the lower panel contains curves broadened by 500 eV. Even in the
latter, admittedly extreme case, the PWFFA offset and subsequent overestimation
of the high-$\omega$ tail is still quite prominent. The \textit{ad hoc} $f$-sum
scaling results in a tail that is only slightly overestimated, but at the cost
of an extreme deficit beneath the peak.

At the higher densities typical of WDM, the core contribution becomes even more
important. As the density is increased, the Fermi level increases relative to
the bottom of the valence band resulting in a broader valence contribution.
The core wavefunction, on the other hand, is only weakly dependent on density
(at least for modest compression, where the cores from neighboring sites have
negligible overlap). Subsequently, as shown in Fig.~\ref{fig:compressed}
(\textit{cf.} Fig.~\ref{fig:expt}, top panel), the core contribution is
relatively larger in the peak region, increasing the importance of a
numerically accurate theoretical treatment.

\subsection{Implications}
\label{sect:implications}

We now turn to implications of an incorrect core treatment on the
interpretation of XRTS spectra from WDM\@.  It is important to recognize the
difficulty of these experiments and their analysis.  The intrinsic width of
backlighter x-ray sources fundamentally limits the energy resolution obtainable
in this measurement technique.  The low flux and need for single-shot
measurement requires the use of low-resolution spectrometers with limited
spectral range further decreasing the resolution while also complicating background
characterization.  This uncertainty in the background subtraction makes $f$-sum
normalization especially difficult, and thus the spectra often can not be
reliably placed into absolute units.  Furthermore, the highest likelihood
background is necessarily dependent upon assumptions made about the core
contribution to the spectrum.

The complicated interplay between the various degrees of freedom present in
such fits makes it difficult to state the exact implications of using the
PWFFA in the extraction of thermodynamic state variables in published
work.\pwffacites It is likely that, in order for a good fit in the
high-$\omega$ tail to be obtained, the ionization state must be overestimated.
This could explain the discrepancy noted by Fortmann, et
al.,\cite{Fortmann2012} between their best-fit ionization state found using
the PWFFA and earlier work that appears to use the
HM.\cite{Lee2009,Glenzer2010} Beyond that, the net effect on extracted
thermodynamic parameters is unclear.  However, due to this previously
undiagnosed systematic uncertainty, re-evaluation of existing experimental
data\pwffacites using a more appropriate core calculation and a maximum
likelihood treatment of the background is necessary.

\begin{figure}
  \includegraphics{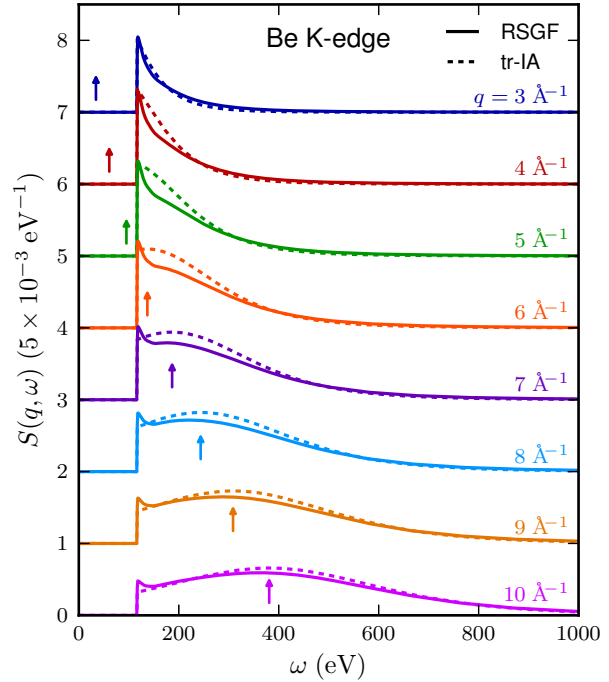}
  \caption{(Color online.) Comparison of theoretical core contribution to Be $K$-edge XRTS calculated using RSGF and truncated IA methods as a function of momentum transfer. Vertical arrows are located at the free-particle Compton shifts.}
  \label{fig:be_vs_q}
\end{figure}

\begin{figure}
  \includegraphics{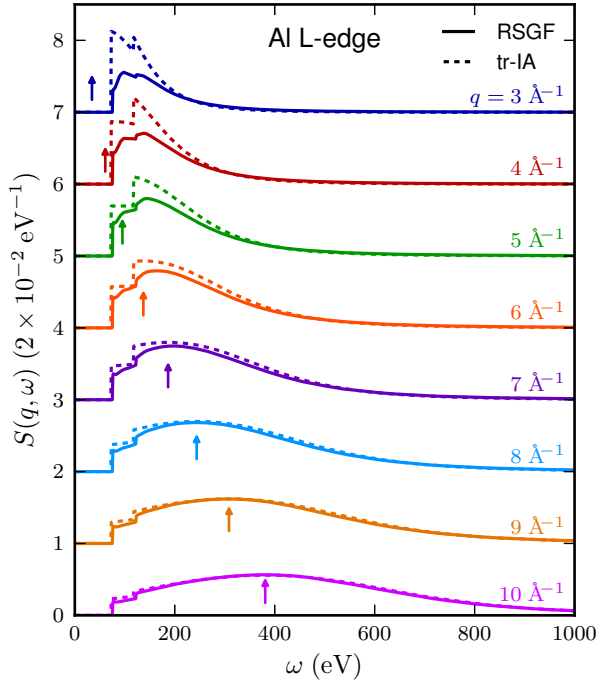}
  \caption{(Color online.) Same as Fig.~\ref{fig:be_vs_q}, except for Al $L_1$- and $L_{2,3}$-edge XRTS\@.}
  \label{fig:al_vs_q}
\end{figure}

\begin{figure}
  \includegraphics{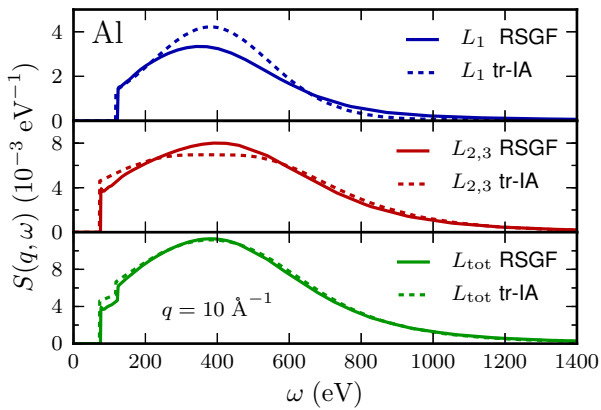}
  \caption{(Color online.) Comparison of RSGF and IA calculations for $L_1$ ($2s$) and $L_{2,3}$ ($2p$) subshells, along with the combined spectra at high $q$. While the RSGF and IA differ for individual subshells, agreement is recovered for the combined spectra. }
  \label{fig:al_by_edge}
\end{figure}

\subsection{Future Practice}
\label{sect:future_practice}

The limitations of the PWFFA bring up the question of best future practice for
fitting the core-shell XRTS from WDM\@.  This will become particularly
important when higher resolution WDM-XRTS experiments are performed at the
Linac Coherent Light Source and the National Ignition Facility.  If such
experiments are to reach their full scientific potential, errors on the scale
of those given by the PWFFA must be avoided.  An ideal treatment
would include self-consistent determination of occupied and unoccupied
electronic states including condensed-phase effects.  Also, the decrease of
ionization potentials with increased density (i.e., \textit{continuum
lowering}) should be either implicitly present in the calculation, or tunable
using models from plasma physics.\cite{Ecker1963,Stewart1966,Zimmerman1980}

Of the methods we have presented, the RSGF calculation most closely describes
the ambient experimental data.  Condensed-phase effects are included explicitly in
final states.  Although a frozen atomic core wavefunction is
used, this is a common feature of all techniques under consideration, and
should be sufficient at modest densities where core overlap is expected to be
negligible.  The primary limitation of the RSGF approach is that it is currently
unclear how to incorporate continuum-lowering effects.

Alternatively, one can use an \textit{ad hoc} modification of the low-energy-transfer
tail of the IA
\begin{equation}
  S_{\rm tr-IA}(q,\omega) = S(q,\omega) \left(1 - \frac{1}{e^{\beta(\omega-E_B)} + 1}\right).
  \label{eq:S_tr-IA}
\end{equation}
We will refer to this approach, which for $T=0$ simply truncates the spectrum below binding energy,
as the \textit{truncated IA} (tr-IA).
This allows straightforward application of 
continuum-lowering models to adjust the binding energy.  A similar
approach, using screened hydrogenic wavefunctions for the initial state instead
of Dirac-Fock is discussed in Gregori, et al.\cite{Gregori2004}, but only in the
context of smaller $q$, where the core contribution is relatively small.  In
Figs.~\ref{fig:be_vs_q},~\ref{fig:al_vs_q},~and~\ref{fig:al_by_edge}, we explore the accuracy and
applicability of the tr-IA  at T=0 by comparing it with RSGF calculations for a range
of momentum transfers for the $K$ shell of Be and $L$ shell of Al.  Note that all
calculations are in absolute units.

As long as the free-particle Compton shift (shown by vertical arrows) is a few
times the edge energy, the tr-IA is in reasonable agreement with the RSGF
calculation.  Since the IA satisfies the $f$-sum rule by construction, the
truncation of the low $\omega$ tail results in slight $f$-sum violations
($\lesssim 5\%$ for Al, $q \ge 6$ \AA$^{-1}$).  We also note that for the Al
$L$ shell, the tr-IA and RSGF calculations differ for the individual subshells
(Fig.~\ref{fig:al_by_edge}, upper two panels). However, agreement is recovered
after combining to form the total $L$-shell contribution
(Fig.~\ref{fig:al_by_edge} lower panel).

Recently, another approach to modeling XRTS from WDM has been discussed by
Johnson, et al.\cite{Johnson2012}  They use an \textit{average-atom} model,
which gives a significant improvement in the treatment of the free-electrons
compared to a simple jellium model.  Unfortunately, the average-atom binding
energies disagree significantly with experiment, limiting the accuracy of the
bound-free contribution to the XRTS spectrum. Johnson, et al.\cite{Johnson2012}
also consider applying the PWFFA to the average-atom $1s$ state for Be and find
similar qualitative discrepancies as we have discussed here.

In summary, for XRTS experiments with modest energy resolution at high
momentum transfer it should be sufficient to treat the bound-free contribution
with a truncated IA, where the truncation energy is adjusted to include the
effects of continuum lowering.  However, the IA ceases to be accurate at lower
momentum transfers. If continuum-lowering shifts and temperatures are
negligible compared to the desired energy resolution, then the RSGF approach
can be immediately applied at any momentum transfer. Further investigation is
needed to determine if continuum lowering can be calculated or included
empirically within the RSGF framework.

\section{Conclusion}
\label{sect:conclusions}

We have discussed several techniques for calculating the core-shell
contribution to XRTS and compared with experimental data collected from
polycrystalline Be under ambient conditions.  Of the techniques considered,
the real-space Green's function method best describes the data.  However, a
simple \textit{ad hoc} truncation of the impulse approximation is reasonably
accurate at higher momentum transfers and allows more straightforward
inclusion of continuum lower effects.  The accuracy of this truncated IA as a
function of $q$ has been explored by comparing with RSGF calculations for both
the Be $K$-shell and Al $L$-shell.  On the other hand, the plane-wave
form-factor approximation, which has been used in the interpretation of
several WDM experiments\pwffacites is quantitatively and qualitatively
inaccurate due to an inconsistent treatment of the single-particle
Hamiltonian.  Re-evaluation of the experimental data using a more accurate
core calculation and maximum likelihood background subtraction is recommended.
More importantly, an accurate treatment of the bound-free XRTS from WDM will
be necessary when higher resolution experiments are performed at the Linac
Coherent Light Source and the National Ignition Facility.  We believe we have
also motivated the need for cross-method comparisons and the usefulness of
exchange of both theoretical and experimental techniques between the
condensed-matter and dense-plasma communities.

\begin{acknowledgments}

This work was supported by the US Department of Energy, Office of Science,
Fusion Energy Sciences and the National Nuclear Security Administration,
through grant DE-SC0008580.  We thank C. Fortmann, S. Glenzer, G. Gregori, T.
Doeppner, J. Rehr, J. Kas, F. Vila and D. Riley for many enlightening
discussions.

\end{acknowledgments}

\end{document}